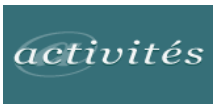

# Étude de l'utilisation d'un environnement numérique de formation : méthode de remise en situation à l'aide de traces numériques de l'activité

*Studying how trainee teachers use an online learning environment: resituating interviews supported by digital traces*

Simon Flandin, Marine Auby et Luc Ria

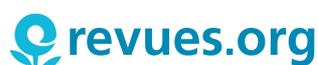

Ce document a été généré automatiquement le 18 octobre 2016.

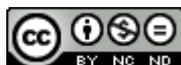

# Étude de l'utilisation d'un environnement numérique de formation : méthode de remise en situation à l'aide de traces numériques de l'activité

*Studying how trainee teachers use an online learning environment: resituating interviews supported by digital traces*

**Simon Flandin, Marine Auby et Luc Ria**

## NOTE DE L'ÉDITEUR

1   Cette contribution rend compte d'une étude conduite dans le programme empirique et technologique du « cours d'action » (Poizat, Durand, & Theureau, sous presse ; Theureau, 2006, 2009), un programme d'anthropologie cognitive basé sur trois hypothèses théoriques fondamentales : (i) l'hypothèse de *l'enaction*, c'est-à-dire d'un couplage structurel dynamique entre l'acteur et son environnement (Varela, 1989), (ii) l'hypothèse de *la conscience préréflexive*, tenue pour l'effet de surface de cette dynamique, à laquelle l'acteur a accès moyennant certaines conditions méthodologiques et (iii) l'hypothèse de *l'activité-signe*, selon laquelle l'activité est une sémiose continue et peut être décrite comme une concaténation de *signes*. Ce programme articule de façon organique la production de savoirs sur les activités humaines et la conception d'aides au développement des acteurs (Durand, 2008 ; Leblanc, Ria, Dieumegard, Serres, & Durand, 2008).

2    Notre étude s'inscrit plus précisément dans un volet de ce programme consacré à la formation des enseignants débutants, caractérisé sur le plan technologique par (i) l'analyse de leur travail ordinaire, notamment en contexte d'intervention difficile (e.g. Ria, 2006) ; (ii) l'identification de certaines caractéristiques typiques et critiques de ce travail à prioriser en formation (e.g. Ria & Rayou, 2008) ; (iii) la modélisation de trajectoires professionnelles jalonnées par des « passages à risque » et devant être accompagnées (Ria, 2009, 2012) ; (iv) la conception de ressources innovantes, notamment vidéo, pour instrumenter cet accompagnement (Ria, Serres, & Leblanc, 2010) ; (v) la conception d'un environnement numérique de formation (ENF par la suite) et son implémentation en ligne[1] (Leblanc & Ria, 2014 ; Ria, & Leblanc, 2011) ; (vi) l'étude de l'utilisation de cet ENF dans différents contextes, configurations et modalités (Flandin, Auby & Ria, accepté ; Flandin, Leblanc & Muller, 2015 ; Leblanc & Sève, 2012 ; Lussi Borer & Muller, 2014) ; (vii) l'étude de l'influence de ces utilisations sur l'activité des enseignants en classe (Flandin & Ria, 2014b ; Leblanc, 2014) ; (viii) la « reconception » itérative de l'ENF par différents aménagements et développements, et la proposition de nouveaux principes de conception (Flandin, 2015 ; Flandin & Ria, 2014a ; Flandin, Ria & Picard, 2015).

3    Le présent article rend compte d'une étude « pilote » menée dans le cadre d'un travail doctoral (Flandin, 2015) portant sur la vidéoformation des enseignants stagiaires, et contribuant (i) à une meilleure compréhension des expériences s'accompagnant d'apprentissage et des principes de conception susceptibles de les favoriser et (ii) à un processus de conception continuée de l'ENF précité et au développement de nouveaux outils et dispositifs. Ce travail doctoral s'est intéressé en particulier à *la modalité d'autonomie* — relativement à l'utilisation de l'ENF — et donc, notamment, à la façon dont étaient consultées (ou non) les différentes ressources vidéo et textuelles. La première question de recherche était : « Comment l'activité des enseignants stagiaires s'organise-t-elle en situation d'utilisation autonome de l'ENF ? » Elle a nécessité le développement d'une méthode d'analyse de l'activité des stagiaires en situation d'utilisation de l'ENF qui perturbe le moins possible cette activité, de façon à simuler de manière vraisemblable une utilisation autonome telle qu'elle pourrait exister, hors de contexte ordinaire, hors de la recherche. En nous basant notamment sur les travaux menés par Leblanc (2001, 2012) sur l'autoformation « hypermédia » de cadres sportifs, nous avons ainsi conçu une étude pilote visant à (i) comprendre et modéliser l'organisation locale de l'activité d'un utilisateur (enseignant) en situation d'utilisation de l'ENF, en cherchant des similarités et récurrences dans son cours d'action (ii) dans la perspective de contribuer, dans les études suivantes, à la compréhension de l'activité d'autres utilisateurs par comparaison, assimilation et différenciation des modélisations produites et (iii) afin d'alimenter, *in fine*, un processus de conception continuée par l'identification de leviers possibles d'amélioration de l'ENF. Ce texte en rend compte en mettant en exergue les options méthodologiques retenues et en les justifiant.

# 1. Méthode de recueil de données

4    La conception d'un observatoire ad hoc implique des choix de méthode maximisant la richesse et l'exploitabilité des matériaux relativement aux objets considérés. Ce principe conduit souvent à des compromis privilégiant certaines dimensions de l'activité (Theureau, 2006). Le premier choix que nous avons effectué concerne l'entretien de remise en situation de l'acteur (noté ERS par la suite), différé de l'activité en train de se faire, à partir de traces numériques de cette activité. Nous avons préféré l'ERS aux verbalisations simultanées et interruptives qui tendent à « ruiner » l'activité par les perturbations qu'elles constituent (Theureau, 2010), et qui auraient été incompatibles avec la simulation d'une autonomie de l'utilisateur.

## 1.1. Les données d'observation de l'activité en situation d'utilisation de l'ENF

5    Après un rappel du déroulement prévu, l'utilisateur était laissé seul dans la pièce, en situation d'utilisation autonome de l'ENF (voir les détails dans la partie 3.2.). Il lui était demandé d'utiliser son ordinateur personnel de façon à (i) limiter d'éventuelles perturbations dues au fait d'utiliser une configuration informatique inconnue ; (ii) disposer sensiblement des mêmes possibilités (techniques, organisationnelles) qu'en situation ordinaire, hors de la recherche. L'écran était enregistré en continu à l'aide de l'application *FastStone Capture*[2], qui concatène des captures à un intervalle de temps optimal permettant d'obtenir une impression dynamique très satisfaisante sans requérir trop de mémoire de calcul et de stockage (10 images par seconde). En d'autres termes, l'enregistrement produit était proche de l'écran vu par l'utilisateur au moment de l'utilisation, avec en plus une aide au repérage des actions effectuées à l'aide du pointeur (Figure 1).

Figure 1 : Extrait d'un enregistrement de l'écran au cours d'une situation d'utilisation de l'ENF.
*Figure 1: Sample of a screen capture in a situation of DLE use*

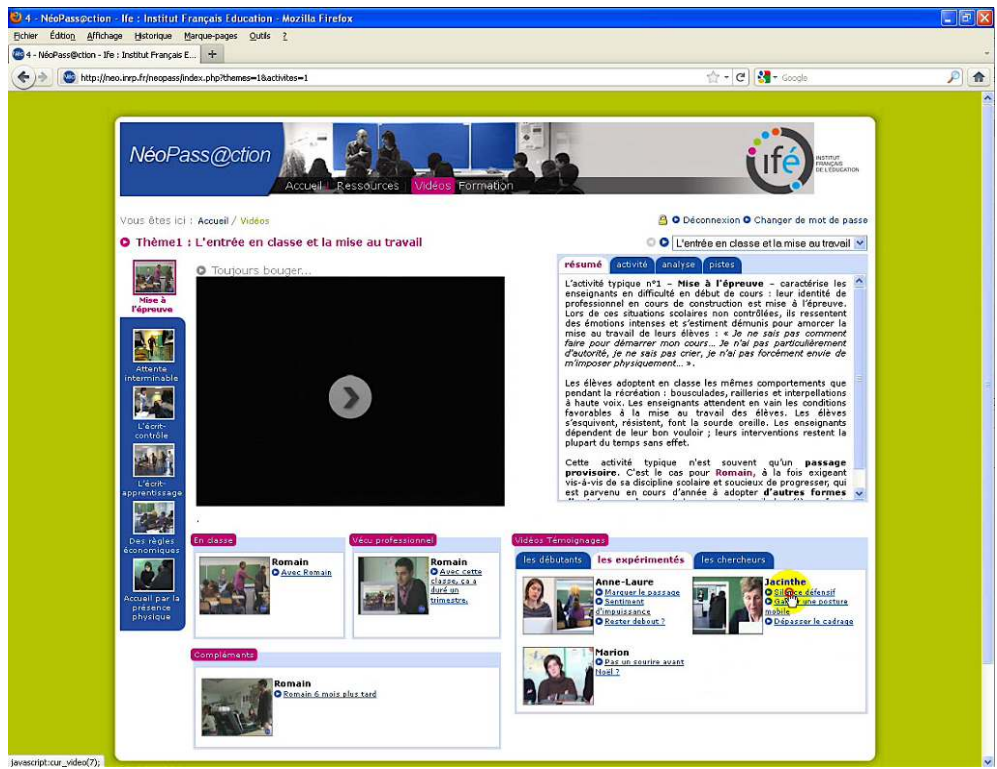

6    On peut remarquer que le pointeur, situé ici dans le coin inférieur droit de l'écran, est entouré d'un halo jaune. De plus, un cercle rouge apparait lorsque l'utilisateur clique. Ces aides visuelles discrètes facilitent le repérage des actions de déplacement et d'interaction avec l'interface sans pour autant contraindre l'attention volontaire à d'autres éléments que le pointeur. À l'issue de la session, l'enregistrement était donc directement exploitable (i) pour être utilisé comme support d'entretien, quelques minutes après la fin de la session de formation, et (ii) pour sa description extrinsèque sous forme tabulaire et chronologique, effectuée lors du traitement des données. La durée maximale contractuellement impartie pour la session était de 45 minutes. À titre indicatif, le temps cumulé de visionnement vidéo possible dans l'ENF utilisé était de 320 minutes, soit quatre heures.

## 1.2. Les données d'entretien de remise en situation (ERS)

7    Le premier enjeu des ERS est de parvenir à « dé-situer » l'acteur (i) de sa situation présente, inhabituelle, adressée à un chercheur, et (ii) des situations d'expression verbale auxquelles il est habitué (oral de concours par exemple dans le cas d'un enseignant stagiaire), pour le « re-situer » dans l'activité à documenter, dont il est confronté aux traces (Theureau, 2010). Les ERS ont été enregistrés à l'aide d'une caméra numérique positionnée sur un pied fixe et d'un micro haute fréquence sans fil. Pour enregistrer à la fois les comportements de l'acteur et les traces d'activité commentées à chaque instant, un moniteur a été ajouté, dupliquant celui utilisé par l'acteur pour visionner les traces de son activité (Figure 2).

8    Avant chaque entretien, le chercheur décrivait à l'acteur le déroulement prévu ; par exemple :

> « Si tu en es d'accord, on va regarder l'enregistrement. Le but est que tu parviennes à décrire ton activité telle que tu l'as vécue, de façon très précise. On se laisse tous les deux la possibilité d'arrêter la vidéo pour en commenter un moment particulier. N'hésite surtout pas à le faire si quelque chose était important pour toi, pour une raison ou une autre, au moment où cela s'est déroulé. Je vais parfois poser des questions, qui peuvent sembler bêtes ou très naïves, mais c'est pour être bien sûr de comprendre. »

9    Au cours de l'entretien, les interventions du chercheur consistaient majoritairement à répéter une fin de phrase pour relancer le commentaire, ou à questionner l'acteur sur une dimension particulière de son expérience.

Figure 2 : Extrait d'enregistrement d'un entretien de remise en situation à l'aide de traces numériques de l'activité.
*Figure 2: Sample of a recording of a resituating interview supported by digital traces of the activity*

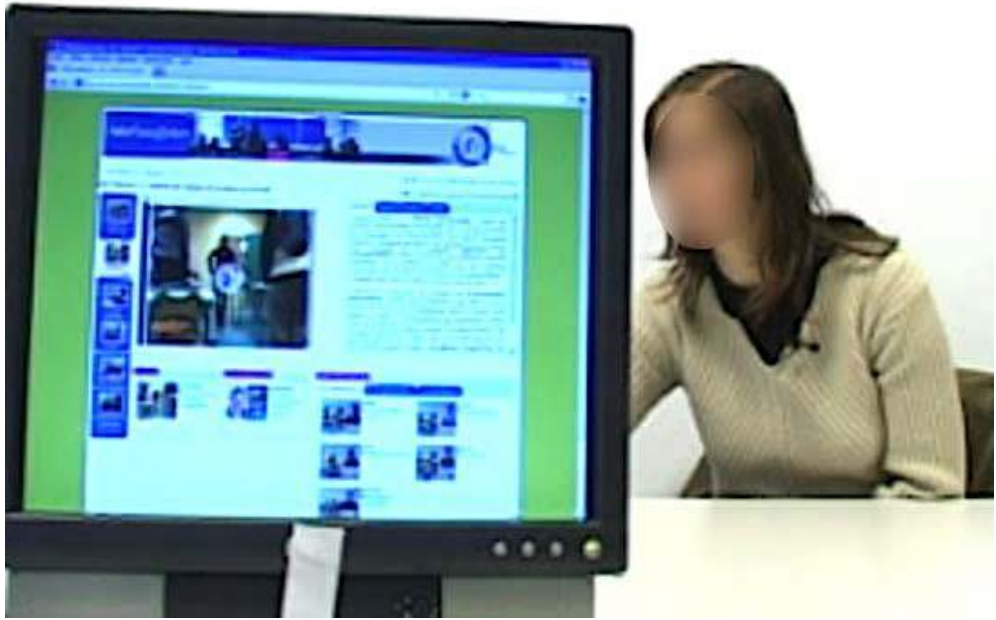

10   L'enregistrement continu de l'écran a constitué une trace numérique de l'activité en situation d'utilisation de l'ENF propice à un entretien de remise en situation. Il limitait très fortement la perte de détails liée aux difficultés de rappel de l'utilisateur, et de possibles affabulations involontaires. L'enregistrement était diffusé sans le son pour deux raisons : (i) permettre un commentaire simultané limitant la durée totale d'entretien (contractuellement contrainte à 60 minutes) ; (ii) encourager l'activité préréflexive au détriment de l'activité réflexive. En effet, revivre « en extériorité » l'utilisation de l'ENF favorise chez l'utilisateur le déploiement d'une activité seconde (une activité *sur* l'activité) que le chercheur devait tenter de limiter autant que possible, afin de resituer l'utilisateur dans l'activité première. Or, s'il était nécessaire qu'il voie les traces visuelles de son action pour la situer et s'y re-situer, il n'était pas forcément nécessaire qu'il en entende les traces sonores. En effet, la durée séparant l'utilisation de l'ENF de l'entretien était courte, ce qui facilitait largement le rappel. Aussi, à de rares exceptions près, l'utilisateur n'avait pas de difficulté à se remémorer les paroles entendues lors de cette utilisation, et en particulier les plus significatives de son point de vue. Néanmoins, le cas échéant, le chercheur et l'utilisateur avaient à tout instant la possibilité de stopper la diffusion pour qu'une action puisse être commentée sans systématiquement obérer la possibilité de commenter les suivantes, dans la limite du temps imparti à l'entretien.

11   Le choix d'un commentaire en partie simultané au déroulement de l'enregistrement résultait donc d'un compromis nécessaire, qui présentait l'avantage de respecter le temps contractuellement imparti à l'entretien, et l'inconvénient de ne pas pouvoir revenir à coup sûr sur l'intégralité de la session d'utilisation de l'ENF, susceptible, de fait, de ne pas être intégralement documentée. La méthode des verbalisations simultanées et interruptives aurait permis de faire l'économie de l'entretien, mais en revanche (i) la présence physique du chercheur dans la situation d'utilisation aurait pu influer fortement sur le comportement de l'utilisateur, qui aurait notamment pu lui prêter des attentes et chercher à s'y conformer ; (ii) les interactions avec le chercheur en cours d'utilisation rompent épisodiquement voire systématiquement la dynamique de l'activité. Il est vrai que cette méthode présente de nombreux avantages (peu couteuse en temps, appareillage léger...) qui la rendent acceptable pour d'autres études soumises à d'autres contraintes, mais pas dans le cas d'une recherche s'efforçant de s'approcher d'une véritable utilisation autonome, telle qu'elle pourrait avoir lieu en dehors de la recherche. Concernant la nature des traces d'activité supports d'entretien, nous avons pu évaluer la pertinence du choix consistant à délaisser les comportements observables de l'utilisateur pour n'enregistrer que ses opérations dans l'interface de l'ENF. Ce choix a eu un inconvénient : nous nous sommes privés de l'expression corporelle (visage, gestes et postures) et donc des rappels qu'elle aurait pu susciter en entretien (tel qu'en autoconfrontation). Cela a eu en revanche deux avantages justifiant l'option prise : (i) une aide performante à la remise en situation dynamique, avec un enregistrement similaire à l'écran vu par l'utilisateur au moment de l'utilisation et (ii) une focalisation accrue sur les interactions avec l'ENF, qui aurait été diminuée par un écran scindé diffusant aussi l'enregistrement du comportement.

## 2. Méthode de traitement des données

12    Les contenus de conscience préréflexive, que l'on cherche à faire expliciter par l'acteur en entretien, sont conçus comme des expériences (ou sémioses), c'est-à-dire des *unités significatives élémentaires* d'activité. L'activité est donc envisagée comme une activité-signe (Theureau, 2006), qui à partir des données d'observation et d'entretien peut être décrite comme une concaténation de signes, et non comme intuition, logique, flux d'information ou suite d'opérations mentales. Ce modèle de description de l'activité se base sur une composition de l'expérience significative en différentes composantes, proposée initialement par Peirce (1931-1935) puis adaptée dans le cadre du cours d'action sous le nom de *signe triadique* (Theureau & Jeffroy, 1994), puis *signe hexadique* (Theureau, 2006). Le recours au *signe hexadique* permet une description de chaque expérience exprimée par l'acteur par sa décomposition en six composantes. Les trois premières constituent la *priméité*, c'est-à-dire l'ensemble des possibles ouverts pour l'acteur par son cours d'action passé jusqu'à cet instant. Elles constituent la structure d'attente de l'acteur (notée E-A-S) :

- *L'engagement* (noté E) est le faisceau de préoccupations (ou d'intérêts immanents) de l'acteur en fonction des actions passées. Ces préoccupations ne se réduisent pas à des buts préétablis prêts à être concrétisés dans l'action, à des productions symboliques explicites et conscientes à tout instant : elles sont de purs possibles, syncrétiques et en partie indéterminés, hérités de l'histoire des couplages passés. L'engagement traduit l'ouverture et la fermeture de ces possibles par l'acteur en situation.
- *L'actualité potentielle* (notée A) est la délimitation des attentes potentielles de l'acteur dans la situation. L'activité étant située dynamiquement, elle est toujours « rétention du tout juste passé » et « protension vers le tout juste à venir », dont l'anticipation correspond à l'actualité potentielle. Elle prolonge concrètement les préoccupations les plus saillantes dans la situation.
- *Le référentiel* (noté S) est l'ensemble des types (ou savoirs), relations entre types (ou dispositions) et principes d'interprétation constituant la culture de l'acteur, et en attente de détermination à chaque instant. Le référentiel est un répertoire d'actions possibles, d'invariants relatifs, non figés et hérités de l'histoire des couplages vécus et « typifiés » (voir plus loin la composante « interprétant »).

13    Les deux composantes suivantes constituent la *secondéité*, c'est-à-dire l'actuel, l'ici et maintenant, le fait, le choc vécu par l'acteur :

- *Le representamen* (noté R) est une perturbation, ce qui « fait signe » pour l'acteur au moment considéré. Il n'est pas une information qui s'impose à l'acteur mais une émergence interdépendante des autres composantes de l'expérience. Il est ce qui se manifeste à la conscience préréflexive de façon perceptive, mnémonique ou proprioceptive. Il sélectionne la fraction de la structure d'attente (E-A-S) qui s'actualise dans la situation, et qui devient alors un objet de connaissance pour la recherche (E-A-S n'étant qu'un ensemble de possibles). La structure actualisée par le representamen est notée eR, aR et sR.
- *L'unité élémentaire* (notée UE) est l'action. Elle est la plus petite fraction d'activité dont un acteur peut rendre compte. Elle est à la fois la résultante des autres composantes du signe et l'expression synthétique de l'activité en cours. En modifiant l'environnement, elle modifie l'ensemble des possibles déjà actualisés par le representamen. Elle peut être une action pratique, une communication, un sentiment, une focalisation ou une interprétation.

14    La sixième et dernière composante, l'*interprétant* (noté I), constitue la *tiercéité*, c'est-à-dire l'élaboration des normes, règles, lois, dispositions prenant une valeur dépassant la situation actuelle. Il s'agit de la mise en relation d'éléments présents et passés, le repérage de régularités, l'évaluation normative, et donc l'apprentissage. Il est conceptualisé en tant que typification (ou typicalisation), c'est-à-dire comme attribution d'une validité (ou d'une invalidité) et d'une certaine valeur de généralité à tout ou partie d'une expérience vécue sur la base d'un gradient de similarité avec des expériences et des circonstances passées (Rosch, 1978 ; Schütz, 1962). Il est une élaboration continue des types et réseaux de types de l'acteur.

15    Le traitement des données a ainsi été effectué à l'aide de l'outillage conceptuel et méthodologique du signe hexadique. Un premier niveau de traitement a consisté à coder et étiqueter les unités significatives de façon à pouvoir les comparer et analyser leur enchainement. Un second niveau de traitement a consisté à repérer des régularités à des fins de modélisation de l'activité.

## 2.1. La documentation des composantes du signe hexadique

16    L'identification et l'étiquetage des composantes des signes hexadiques ont été réalisés à partir d'une mise en relation des données d'observation et des données d'entretien, en accordant le primat aux secondes. Nous présentons ici le questionnement analytique et les conventions d'étiquetage pour renseigner chacune des composantes du signe hexadique (Serres, 2006 ; Trohel, 2005) :

- Pour renseigner le representamen (R) : *quel(s) élément(s) rappelé(s), perçu(s) dans la situation fait (font) signe(s) pour l'utilisateur au moment considéré ?* Le representamen a été étiqueté par un groupe nominal pouvant être suivi d'adjectif(s) qualificatif(s).
- Pour renseigner l'engagement (eR) : *quelle(s) est (sont) la (les) préoccupation(s) saillante(s) dans la situation chez l'utilisateur relativement à ce qui fait signe pour lui au moment considéré ?* Les préoccupations dans la situation ont été étiquetées par un verbe d'action à l'infinitif suivi d'un complément d'objet direct et/ou indirect.
- Pour renseigner l'actualité potentielle (aR) : *quelles sont les attentes concrètes de l'utilisateur dans la situation au moment considéré compte tenu de ses préoccupations dans* la situation ? Quel résultat attend-il de son action ? L'actualité potentielle a été étiquetée par une proposition débutant par « attentes liées à … » suivie des éléments spécifiant ces attentes.

- Pour renseigner le référentiel (sR) : *quels sont les types (savoirs, croyances) mobilisés dans la situation par l'utilisateur pour agir au moment considéré ?* Les types ont été étiquetés à l'aide de propositions comprenant un sujet, un verbe et des arguments complétant le verbe.
- Pour renseigner l'unité élémentaire (UE) : *que fait l'utilisateur au moment considéré ? Que pense-t-il ? Que ressent-il ?* L'unité élémentaire a été étiquetée par un verbe d'action suivi d'un complément d'objet direct et/ou indirect.
- Pour renseigner l'interprétant (I) : *Quels sont les types créés/construits par l'utilisateur au moment considéré ?* Dans ce cas l'interprétant a été étiqueté par une proposition débutant par « Émergence du type… » *Quels sont les types « déjà là » dont la validité augmente pour l'utilisateur au moment considéré ?* Dans ce cas l'interprétant a été étiqueté par une proposition débutant par « Gain de validité du type… » *Quels sont les types invalidés par l'utilisateur au moment considéré ?* Dans ce cas l'interprétant a été étiqueté par une proposition débutant par « Perte de validité du type… »

17    À l'exception de l'unité élémentaire, chaque composante pouvait être renseignée par plusieurs éléments.

## 2.2. La modélisation de l'activité

18    Le cadre d'analyse du cours d'action permet une description détaillée de périodes d'activité singulières, et plus particulièrement une description de la dynamique des couplages acteur-environnement. En dépit de son caractère ouvert et indéterminé, cette dynamique s'organise à différents niveaux en des formes reconnaissables. C'est pourquoi il est possible de parler de « trans-formation » de l'activité même si l'activité est elle-même, intrinsèquement, une dynamique de transformation. Pour repérer ces formes d'activité, il faut disposer d'indices relatifs à des permanences ou des stabilités. Cela se traduit par : (i) la constitution de repères stables ou de possibilités de comparaison entre les phénomènes dans des épisodes d'activité successifs jugés analogues, (ii) la reconstitution, au-delà des spécificités individuelles, de genèses typiques à partir de ces épisodes, (iii) la reconstitution de genèses typiques à partir d'épisodes explorés chez différents acteurs jugés *a posteriori* comme analogues et typiques (adapté de Poizat *et al.*, sous presse.) Ce dernier point est la condition de production d'un savoir anthropologique généralisable à des niveaux de culture dépassant la culture locale des acteurs considérés. Il est bien illustré par exemple par Sève et Leblanc (2003) qui ont décrit, à partir de la comparaison de l'engagement en situation de pongistes et d'utilisateurs d'hypermédia, des séquences d'activité génériques d'exploration et d'exécution[3].

19    Concernant notre recherche, l'étude pilote décrite dans cet article visait le repérage de régularités à différents niveaux :

- Par l'identification du caractère typique de l'expérience de l'utilisateur sur la base du contenu de ses verbalisations : soit l'utilisateur énonçait lui-même le caractère typique de son expérience (en utilisant des formes verbales telles que « à chaque fois », « tout le temps », « systématiquement », etc.) ; soit le chercheur l'identifiait par la mise en évidence d'engagements-types, préoccupations-types, évènements-types, émotions-types ou situations-types à l'échelle d'une utilisation, puis de plusieurs – ce qui a été particulièrement utile à la compréhension de l'expérience-utilisateur ;

- Par l'identification ultérieure de similitudes au niveau des composantes des signes hexadiques entre des analyses provenant de différents utilisateurs ;
- Par l'identification ultérieure de similitudes entre les utilisateurs à différents niveaux d'organisation de l'activité et à différentes échelles temporelles.

20    La modélisation finale des cours d'action des utilisateurs devait permettre une mise en relation entre différents niveaux d'organisation de l'activité des utilisateurs et différentes caractéristiques de la situation d'utilisation de l'ENF, telles que : (i) les caractéristiques propres à la vidéo comme « objet temporel », favorisant la synchronisation et la création d'attentes (Leblanc, 2012) ; (ii) les caractéristiques propres à une « entrée activité » en formation telles que l'authenticité et la typicité des situations (Durand, 2008) ; (iii) les caractéristiques propres à l'ENF, c'est-à-dire être retrouvées à une « pédagogie des trajectoires professionnelles » qui favorise et accompagne les trajectoires des débutants en les modélisant et les didactisant (Durand, 2014 ; Ria & Leblanc, 2011) ; (iv) les caractéristiques propres au dispositif de recherche, c'est-à-dire l'ensemble des limites à l'écologie des situations (Theureau, 2010) ayant des effets transformatifs (planification et organisation des sessions, conduites d'entretien, etc.)

# 3. Mise en œuvre de la méthode

21    Nous avons procédé à l'étude systématique du cours d'action d'un utilisateur en situation d'utilisation de l'ENF. Cette étude pilote avait une double visée : (i) comprendre et modéliser l'organisation locale de l'activité de l'utilisateur et (ii) contribuer ultérieurement à la compréhension de l'activité d'autres utilisateurs par comparaison, assimilation et différenciation des modèles produits. L'analyse a donc consisté en l'élaboration, à partir de l'étude du cours d'action, de catégories d'intelligibilité de différents rangs permettant de rendre compte de l'activité d'un utilisateur lors d'une session. Ces catégories pouvaient ainsi être retrouvées et affinées par l'analyse de l'activité de cet utilisateur lors d'une autre session et par l'analyse de l'activité des autres utilisateurs, ou au contraire abandonnées. Ce processus de stabilisation des catégories est la condition *sine qua non* d'une généralisation des résultats.

22    L'étude pilote concerne la première utilisation de l'ENF par une utilisatrice nommée Anne, la première à participer à la recherche. Nous présentons dans cette partie (i) la description extrinsèque et intrinsèque de son activité ; et (ii) la modélisation de son engagement en unités de différents rangs (séquences, macro-séquences, séries : voir partie 5).

## 3.1. Caractéristiques de la session d'utilisation autonome de l'ENF

23    La session d'utilisation de l'ENF s'est déroulée dans une salle de travail de l'un des deux collèges où travaille Anne, en fin d'après-midi. Elle a duré 47 min. 8 sec. Bien que cette session se soit insérée à la fin d'une journée que l'on peut qualifier de « chargée » (six heures de cours dans deux établissements différents), Anne n'a pas indiqué de fatigue ou autre contrainte contre-indiquant son implication dans l'étude. Elle a utilisé son ordinateur portable personnel.

24    Le protocole lui avait déjà été présenté au cours du recrutement de la cohorte, mais il était important d'effectuer un rappel pour trois raisons principales : (i) lever in extremis d'éventuels malentendus persistants en rappelant les finalités et les étapes caractérisant l'étude ; (ii) contribuer à placer l'utilisatrice dans des conditions favorables à l'activité attendue en rappelant les consignes ; (iii) s'assurer que le contrat d'engagement réciproque entre chercheur et utilisatrice était clair. Avant de commencer la session, le chercheur a donc rappelé rapidement les attentes de l'étude, en formulant les principes suivants :

- tu vas pouvoir utiliser NéoPass@ction, en autonomie, comme si tu t'étais connectée seule, de ta propre initiative ;
- nous sommes convenus d'une durée-cible de 45 minutes, mais tu peux t'arrêter dès que tu as le sentiment que tu l'aurais fait en dehors de cette étude ;
- je n'ai pas d'attentes particulières concernant ce que tu vas faire ou non, et tu as toute liberté pour utiliser ou ne pas utiliser tout ou partie de cet ENF ;
- tout ce qui se passe à l'écran sera enregistré et nous servira plus tard de support d'entretien ;
- l'étude ne vise pas à déterminer si NéoPass@ction est un bon ENF ou non, elle vise à comprendre dans le détail comment les enseignants stagiaires l'utilisent, pour éventuellement, plus tard, élaborer différentes pistes d'amélioration ;
- en cas de problème technique ou autre empêchant l'utilisation de l'ENF, n'hésite pas à m'appeler pour assistance[4].

25    Après ces indications préliminaires, le chercheur a lancé l'application de capture dynamique de l'écran permettant de recueillir les données d'observation, puis a quitté la pièce, ce qui a marqué le début de la session. L'entretien de remise en situation (ERS) qui a suivi a commencé après dix minutes « de pause », et a duré 58 minutes. Il a été conduit selon les modalités décrites précédemment, puis transcrit *verbatim*. À partir de cette étape, il était possible de débuter la construction d'un protocole de description de l'activité à deux volets : un volet de description extrinsèque (données d'observation) et un volet de description intrinsèque (données d'entretien).

## 3.2. Traitement des données d'observation (description extrinsèque)

26    Le codage de l'utilisation de l'ENF a nécessité l'élaboration d'une table combinant : (i) une nomenclature des vidéos de l'ENF (187 items) ; (ii) une nomenclature des éléments non interactifs (tous les éléments visuels non « cliquables » : 40 items) ; et (iii) un index des actions possibles dans l'ENF (3 items). La plateforme étant une interface uniquement graphique, les actions possibles sont celles permises par un dispositif de pointage (pointer, cliquer, scroller[5]). Nous avons ainsi obtenu une table basée sur l'architecture arborescente de l'ENF NéoPass@ction (Figure 3) permettant de coder conjointement les actions de l'utilisatrice et les éléments sur lesquelles elles portaient, soit 681 combinaisons possibles.

Figure 3 : Interface de l'ENF NéoPass@ction telle qu'utilisée par Anne, annotée pour le lecteur (Flandin, 2015).
*Figure 3: Interface of the NeoPass@ction DLE as used by Anne, annotated for the reader (Flandin, 2015)*

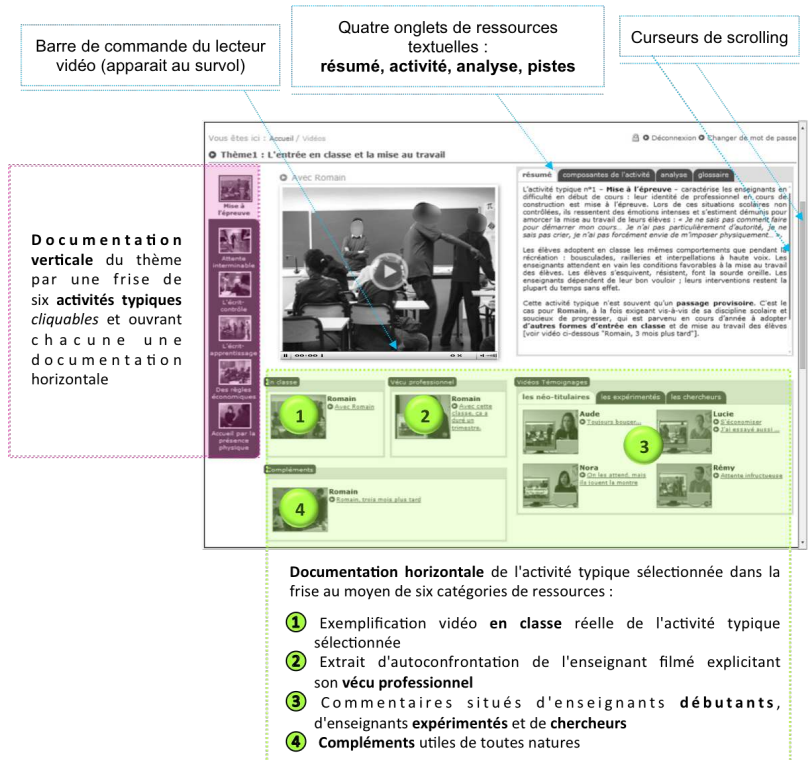

27    La frise verticale à la gauche de l'écran correspond aux six activités différentes typiquement mises en œuvre par les enseignants dans la situation étudiée : la mise au travail des élèves en début de cours (Ria, 2012). Six types d'entretiens vidéo documentent chacune de ces activités typiques (extrait de classe, vécu professionnel de l'enseignant filmé, commentaire de débutants, d'expérimentés, et de chercheurs, compléments divers.)

28    Le codage se déroulait de la façon suivante : l'enregistrement était lu à l'aide du logiciel Kinovéa[6] qui permet de lire et relire des épisodes très courts avec une grande précision (1 % de la vitesse de lecture en temps réel). Chaque action était codée chronologiquement, par écrit, à l'aide de la table. Les données étaient ensuite saisies dans le logiciel Actogram[7], qui a permis leur traitement statistique systématique. Le codage de l'activité manifeste de l'utilisatrice, observable à l'écran, a donc permis d'établir un premier volet du protocole de description.

## 3.3. Traitement des données d'entretien (description intrinsèque)

29    Le second volet a consisté en une transcription *verbatim* des verbalisations recueillies en entretien, mises en correspondance chronologique avec les actions commentées. Ainsi chaque ligne du protocole indique-t-elle un horodatage (un moment de l'utilisation de l'ENF), le codage de l'action produite et son commentaire par l'utilisatrice (lorsque l'action a été commentée en entretien). Le Tableau 1 présente un extrait de ce protocole appliqué au cas d'Anne.

**Tableau 1 : Extrait du protocole de description à deux volets de l'activité d'Anne.**
**Table 1: Sample from a two-part description protocol of Anne's activity**

| Description extrinsèque | | Description intrinsèque |
|---|---|---|
| Horodatage | Actions | **Verbatim (Entretien de remise en situation)** |
| 00:07:17.32 | PV A1 V5 6 | |
| 00:07:18.06 | PV A1 V5 7 *(pointe la vignette codée a1.v5.7)* | Anne : Alors là la vidéo de Marion, c'est le titre qui m'a interpelée. Pour moi "Pas un sourire avant Noël" ça évoquait une position plus froide, avec plus de distance… Je me suis dit "oh là là" c'est peut-être la façon dont elle s'y prend et j'ai été très curieuse. |
| 00:07:18.49 | C A1 V5 7 *(clique sur la vignette codée a1.v5.7)* | Anne : Et quand j'ai vu la vidéo j'étais assez surprise parce que je me suis rendue compte que finalement j'ai pas compris pourquoi ils avaient choisi ce titre. Parce que la phrase a été évoquée une seule fois ; il me semble que ce n'était pas le « thème » général. Il y avait un décalage entre le titre, ce qui a motivé mon envie de voir la vidéo et le contenu. *[Prend une attitude circonspecte.]* On parlait une fois de plus de la mise au travail, de ces choses-là… Et j'ai été surprise. Chercheur : D'accord. Et du coup de quoi elle parlait cette enseignante si elle ne parlait pas de ça ? Anne : Euh… Ben du coup j'ai été tellement surprise par le décalage que j'ai fait un peu l'impasse sur le contenu ! |

30    Dans cet exemple, Anne explicite une expérience « déceptive » qui traduit un décalage entre les attentes générées chez elle par la lecture du titre de la vidéo, et son expérience vécue lors du visionnement. L'analyse de cette séquence (présentée plus loin) va permettre de caractériser ce décalage de façon à produire des savoirs (i) sur l'activité d'Anne et (ii) sur l'ENF appréhendé du point de vue de l'utilisateur.

## 4. Analyse locale du cours d'action : étiquetage des unités significatives et de leurs composantes

31    Trois dimensions complémentaires caractérisent la méthode du cours d'action (Theureau & Jeffroy, 1994) : (i) elle est synthétique pour ce qui relève du découpage/étiquetage des unités significatives ; (ii) elle est analytique pour ce qui relève de leur catégorisation ; (iii) elle est comparative pour ce qui relève de l'identification des invariants, des variations et des facteurs de variation. Cela nécessite des analyses progressives (traitement chronologique) et régressives (traitement typologique) qui précisent et affinent itérativement les modèles produits.

32    L'étiquetage des unités significatives résulte de l'analyse conjointe des deux volets du protocole, vérifiée, au besoin, par la consultation des enregistrements originaux. Il s'agissait à cette étape de décomposer les actions commentées en signes hexadiques selon la méthode déjà décrite. De même que toutes les actions n'ont pas pu être commentées, toutes les composantes des actions n'ont pas pu être documentées. Certaines pouvaient être inférées rétrospectivement, de manière prudente, à l'aide d'une analyse plus globale du cours d'action. Pour le cas d'Anne, cette analyse a abouti à un « récit réduit » de l'activité en situation d'utilisation de l'ENF concaténant 53 unités significatives. L'enchaînement des unités 12 et 13 est documenté dans le Tableau 2 à titre d'exemple. Ces unités sont dites élémentaires lorsqu'elles correspondent précisément à une action, c'est-à-dire à la plus petite fraction d'activité ayant fait l'objet d'un commentaire situé ; elles sont dites non-élémentaires lorsqu'elles correspondent à une fraction significative plus large d'activité, qui dépasse la seule action.

Tableau 2 : Décomposition hexadique et étiquetage des actions documentées dans le Tableau 1.
*Table 2: Hexadic decomposition and labelling of the actions documented in Table 1*

| Composantes du signe | Unités significatives |
|---|---|
| **Representamen (R) :** Le titre « Pas un sourire avant Noël » perçu comme étant prometteur ; un sentiment de curiosité<br>**Engagement (eR) :** Identifier un contenu prometteur ; En savoir plus sur le « cadrage » des élèves ; Consulter l'avis d'un pair expérimenté en vidéo<br>**Actualité potentielle (aR) :** Attentes liées à l'identification d'un titre de vidéo prometteur<br>**Référentiel (sR) :** « Les expérimentés sont les plus à même de proposer des façons de faire intéressantes »<br>**Interprétant (I) :** Émergence du type « L'ENF propose des contenus relatifs au cadrage des élèves » | **UE12**<br>**Observe la vignette de Marion dans l'ENF**<br>**(a1_v5_7)** |
| **Representamen (R) :** Le contenu relatif « à la mise au travail » ; Le décalage entre ses attentes et ce contenu ; Un sentiment de déception<br>**Engagement (eR) :** Trouver des pistes pour un bon cadrage des élèves<br>**Actualité potentielle (aR) :** Attentes liées au visionnement d'une « position enseignante froide et distanciée » ; à l'identification de pistes pour améliorer sa propre intervention<br>**Référentiel (sR) :** Les expérimentés sont les plus à même de proposer des façons de faire intéressantes<br>**Interprétant (I) :** Émergence du type « Des écarts sont possibles dans l'ENF entre ce que le titre laisse à penser et le contenu réel de la vidéo » | **UE13**<br>**Visionne le témoignage de Marion dans l'ENF**<br>**(a1_v5_7)** |

33    L'analyse de l'activité d'Anne montre des attentes particulières suscitées chez elle par la lecture de l'intitulé de la ressource vidéo (« Pas un sourire avant Noël »). Ce titre incite Anne à consulter la ressource car elle semble alors prometteuse vis-à-vis de ses préoccupations, relatives à l'identification de pistes concrètes pour « cadrer » les élèves. Cependant, au cours de la consultation, elle ressent un décalage entre ses attentes et le contenu effectif de la ressource. Cette expérience déceptive s'accompagne de l'élaboration d'un savoir sur les contenus de l'ENF qui, nouvellement intégré au référentiel d'Anne, va influer sur la suite de son utilisation.

34    Cet exemple met donc à jour un décalage entre l'utilité perçue et l'utilité réelle de la ressource a1_v5_7 du point de vue d'Anne. La documentation de ce type de décalage est particulièrement intéressante pour le chercheur (i) car il influe fortement sur la nature de l'expérience-utilisateur et (ii) s'il est documenté plusieurs fois pour une même ressource avec plusieurs utilisateurs, cela peut constituer un motif de reconception. Dans les deux cas, une analyse globale du cours d'action est requise.

# 5. Analyse globale du cours d'action : identification des relations de cohérence et construction des graphes

35    Après la construction du protocole à deux volets, le niveau d'analyse locale constitue une première étape mais ne suffit pas pour comprendre les cohérences[8] d'ensemble dans l'organisation de l'activité. L'analyse globale a ensuite consisté à identifier des unités significatives de rang supérieur aux unités du récit réduit en établissant des relations de cohérence entre elles, c'est-à-dire un engagement identique de l'utilisatrice à différents moments de la session. Ces relations de cohérence sont de trois rangs distincts (Theureau & Jeffroy, 1994) : (i) séquences (enchaînements d'unités élémentaires) ; (ii) macro-séquences (compositions de séquences relevant d'un même engagement de l'utilisateur dans le temps) ; et (iii) séries (engagements globaux à l'échelle de la pratique dans laquelle est engagé l'utilisateur). Elles permettent une comparaison de différents niveaux d'organisation de l'activité au sein d'une même session d'utilisation de l'ENF, mais aussi entre deux utilisations d'un même utilisateur ou entre plusieurs utilisateurs.

## 5.1. Identification et étiquetage des séquences

36 Nous avons procédé à un second étiquetage visant à identifier des relations séquentielles entre les unités significatives. Ces relations peuvent être continues lorsque les unités s'enchaînent (par exemple, U15, U16 et U17 relèvent d'une même séquence) ou discontinues dans le cas contraire (comme U3, U7 et U12 — voir Tableau 3). Les unités significatives caractérisant le même engagement-type de l'utilisatrice en situation ont ainsi été étiquetées dans une même séquence. À ce niveau de description de l'activité, il devient possible d'identifier des engagements typiques dans l'activité en situation d'utilisation de l'ENF.

Tableau 3 : Catégorisation d'unités significatives entretenant une relation séquentielle discontinue (séquence 4).
*Table 3: Categorization of meaningful units linked in a sequential and discontinuous fashion*

| Unités significatives et composantes du signe | Séquence (S) |
|---|---|
| *UE 3 : Observe la vignette de Aude dans l'ENF (a1_v4_5)*<br>**Representamen (R)** : Le titre prometteur « Toujours bouger » évoquant la mobilité ; Un sentiment de curiosité<br>**Engagement (eR)** : Identifier une discussion prometteuse de l'intervention en classe de Romain<br>**Actualité potentielle (aR)** : Attentes liées à l'identification d'un titre de vidéo prometteur<br>**Référentiel (sR)** : L'intervention en classe de Romain manque de mobilité<br>**Interprétant (I)** : Émergence du type « L'ENF propose des contenus relatifs à la mobilité » | |
| [UE4, UE5, UE6] | |
| *UE 7 : Observe la vignette de Jacinthe dans l'ENF (a1_v5_4)*<br>**Representamen (R)** : L'âge apparent de l'enseignante, signe d'expérience ; Le titre prometteur « Garder une posture mobile » évoquant la mobilité<br>**Engagement (eR)** : Identifier une discussion prometteuse de l'intervention en classe de Romain ; Comparer la discussion par une débutante à la discussion par une expérimentée<br>**Actualité potentielle (aR)** : Attentes liées à l'identification d'un titre de vidéo prometteur<br>**Référentiel (sR)** : L'intervention en classe de Romain manque de mobilité<br>**Interprétant (I)** : Gain de validité du type « l'ENF propose des contenus relatifs à la mobilité » | **S4**<br>**Identifier un contenu prometteur** |
| [UE8, UE9, UE10, UE11] | |
| *UE12 : Observe la vignette de Marion (a1_v5_7)*<br>**Representamen (R)** : Le titre prometteur « Pas un sourire avant Noël », évoquant le cadrage des élèves ; Un sentiment de curiosité<br>**Engagement (eR)** : Identifier un contenu prometteur concernant la question du cadrage des élèves ; Consulter l'avis d'un pair expérimenté<br>**Actualité potentielle (aR)** : Attentes liées à l'identification d'un titre de vidéo prometteur<br>**Référentiel (sR)** : Les expérimentés sont les plus à même de proposer des façons de faire intéressantes<br>**Interprétant (I)** : Émergence du type « L'ENF propose des contenus relatifs au cadrage des élèves » | |

37 Les trois unités significatives décrites dans le Tableau 3 constituent une même séquence au cours de laquelle l'activité d'Anne consiste en une exploration de l'interface de l'ENF, orientée vers l'identification de contenus prometteurs vis-à-vis de ses préoccupations et intérêts pratiques.

## 5.2. Séquences prospectives et rétrospectives

38 Une séquence est dite prospective lorsqu'il y a un lien, une correspondance entre les unités successives qui la composent, du point de vue de l'utilisateur et de son engagement au moment considéré. En effet, l'expérience vécue dans la première unité ouvre une structure d'attentes déterminée (E-A-S), et donc notamment un résultat attendu lorsque l'action (UE) est censée, du point de vue de l'utilisateur, s'accompagner d'effets perceptibles. Lorsque l'un ou plusieurs des representamens (R) de l'unité suivante correspond(ent) à cette attente, la séquence est dite prospective. C'est le cas dans l'exemple du Tableau 2 lorsque Anne, observant les vignettes (UE12), a des attentes relatives à l'identification d'un contenu prometteur sur le thème du « cadrage des élèves », et qu'elle identifie le titre « Pas un sourire avant Noël », representamen qui correspond à ces attentes. Par conséquent, l'action suivante d'Anne consiste à cliquer sur la vidéo pour la consulter (UE13), ce qui prolonge cette séquence prospective (étiquetée S4. Identifier un contenu prometteur), et s'accompagne d'un apprentissage sur l'ENF (élaboration du type « L'ENF propose des contenus prometteurs relatifs au cadrage/ contrôle des élèves »).

39 En revanche, lorsque le résultat attendu n'émerge pas ou lorsqu'un résultat inattendu émerge, la séquence est dite rétrospective. Si nous poursuivons notre exemple, lors du visionnage de la vidéo (UE13), Anne a le sentiment d'un décalage entre ses attentes et le contenu, qui s'accompagne également d'un apprentissage sur le dispositif (diminution de la validité du type « L'ENF propose des contenus prometteurs relatifs au cadrage des élèves », élaboration du type « Des écarts sont possibles entre ce que le titre laisse penser et ce que la vidéo propose »). Toute séquence rétrospective ne s'accompagne pas d'un retentissement négatif ou déceptif (l'utilisateur peut potentiellement avoir « une bonne surprise », par exemple) mais généralement, une expérience-utilisateur positive requiert un sentiment régulier de prise et de maitrise de l'interface (Amadieu & Tricot, 2006). Aussi toutes les séquences rétrospectives ont-elles été systématiquement analysées par le chercheur comme autant de leviers possibles d'amélioration de l'ENF. L'identification de ces séquences prospectives et rétrospectives peut en effet renseigner le concepteur à deux titres : (i) la connaissance de l'expérience-utilisateur : l'expérience plutôt positive d'une utilisation qui se déroule en fonction de choix délibérés *versus* l'expérience plutôt négative d'un déroulement plus ou moins indépendant de ceux-ci ; (ii) la conception continuée : la mise en perspective, au niveau local, entre l'activité attendue par les concepteurs et l'activité réelle et donc l'étude du rapport élémentaire entre logique de conception et activités d'utilisation à des fins d'aménagement-reconception.

## 5.3. Identification et étiquetage des macro-séquences et des séries

40 L'analyse globale du cours d'action nous a permis un troisième étiquetage : celui des séquences entretenant, dans le cadre d'un thème plus large, des relations de cohérence macro-séquentielle. Quatre macro-séquences ont ainsi été identifiées, et ont révélé après analyse de l'ensemble du corpus de données quatre engagements-types cohérents dans l'activité d'Anne :

1. MS1. Organiser et optimiser sa session de formation
2. MS2. Suivre et élaborer une cohérence d'utilisation

3. MS3. Sélectionner des objets pertinents

4. MS4. Trouver des pistes pour l'intervention

41    Deux d'entre elles, à dominante exécutoire, visaient plutôt à avancer efficacement dans la consultation des ressources et dans les apprentissages ; les deux autres, à dominante exploratoire, visaient plutôt à préparer, optimiser par anticipation l'utilisation de l'ENF et les apprentissages afférents. On retrouve régulièrement les quatre macro-séquences du début à la fin de la session d'Anne.

42    Enfin, un quatrième et dernier étiquetage, celui des relations de cohérence sérielle, nous a permis de réduire les multiples engagements initialement repérés dans la description de l'activité de l'utilisatrice à une alternance binaire de séries :

1. Sé1. Investir les possibles du dispositif

2. Sé2. Développer son activité de travail

43    Des unités significatives entretiennent une relation sérielle lorsque l'engagement global de l'utilisateur émerge à l'identique dans ces différentes unités indépendamment de leur dynamique d'engendrement. Les séries ont donc été identifiées à partir du repérage : (i) des unités significatives redondantes au cours de la session ; et (ii) des préoccupations globales correspondantes, parfois exprimées par l'utilisatrice au cours de l'entretien.

44    L'analyse globale du cours d'action d'Anne nous a donc conduit in fine à catégoriser les 53 unités significatives issues de l'analyse locale en 22 séquences, quatre macro-séquences et deux séries (Tableau 4). La numérotation des séquences est chronologique : elles ont été numérotées par ordre d'occurrence dans le cours d'action. Leur organisation dans le tableau est en revanche typologique : elles sont mises en correspondance avec la macro-séquence et la série auxquelles elles appartiennent.

Tableau 4 : Enchâssement des séquences, macro-séquences et séries organisant l'utilisation de l'ENF par Anne.
*Table 4: Sequential, macro-sequential and serial embedding, organizing Anne's use of the DLE*

| Séquences (S) | Macro-séquences (MS) | Séries |
|---|---|---|
| S1. Déterminer par où commencer | MS1. Organiser et optimiser sa session de formation (*MS à dominante **exécutive** relative aux contraintes et ressources de la **scénarisation ouverte**) | Sé1. Investir les possibles du dispositif |
| S6. Privilégier le type de contenu le plus pertinent | | |
| S11. Avancer dans la consultation des ressources | | |
| S2. Consulter le témoignage d'un pair débutant | | |
| S3. Suivre la progression « horizontale » | | |
| S7. Consulter le témoignage d'un pair expérimenté | MS2. Suivre et élaborer une cohérence d'utilisation (*MS à dominante **exécutive** relative aux contraintes et affordances **de l'interface**) | |
| S15. Suivre la progression « verticale » | | |
| S16. Prendre connaissance de la situation initiale | | |
| S19. Consulter le témoignage d'un pair déjà consulté | | |
| S4. Identifier un contenu prometteur sur un thème défini | MS3. Sélectionner des objets pertinents (*MS à dominante **exploratoire de l'interface** relative à un intérêt pratique*) | Sé2. Développer son activité de travail |
| S12. Identifier un contenu prometteur sans thème défini | | |
| S13. Identifier un contenu prometteur relativement à un problème personnel | | |
| S5. Trouver des pistes relatives à la mobilité | MS4. Trouver des pistes pour l'intervention (*MS à dominante **exploratoire de l'activité enseignante** relative à un intérêt pratique*) | |
| S8. Trouver des pistes relatives aux stratégies de défense | | |
| S9. Trouver des pistes relatives au cadrage des élèves et aux stratégies de défense | | |
| S10. Trouver des pistes relatives au cadrage des élèves | | |
| S14. Trouver des pistes relatives aux passages à risque | | |
| S17. Comprendre les modes de développement de l'activité | | |
| S18. Investiguer une situation dans laquelle elle se reconnaît | | |
| S20. Trouver des pistes relatives à l'économie de paroles | | |
| S21. Trouver des pistes relatives à la présence en classe | | |
| S22. Trouver des éléments de synthèse sur l'objet de formation | | |

## 5.4. Construction des graphes (ou chronogrammes) de l'activité en situation d'utilisation de l'ENF

45    Au terme du traitement de ces données, il a été possible de modéliser l'activité d'Anne sous forme de deux graphes : l'un modélisant son organisation d'un point de vue extrinsèque, c'est-à-dire la chronique de la partie observable des interactions utilisatrice/ ENF (Figure 4) ; et l'autre modélisant son organisation d'un point de vue intrinsèque, c'est-à-dire la chronique de la partie subjective des interactions utilisatrice/ENF. La Figure 5 illustre les cinq premières minutes de ce graphe intrinsèque.

Figure 4 : Extrait du graphe de description extrinsèque de l'activité d'Anne.
*Figure 4: Extract from the extrinsic description graph of Anne's activity*

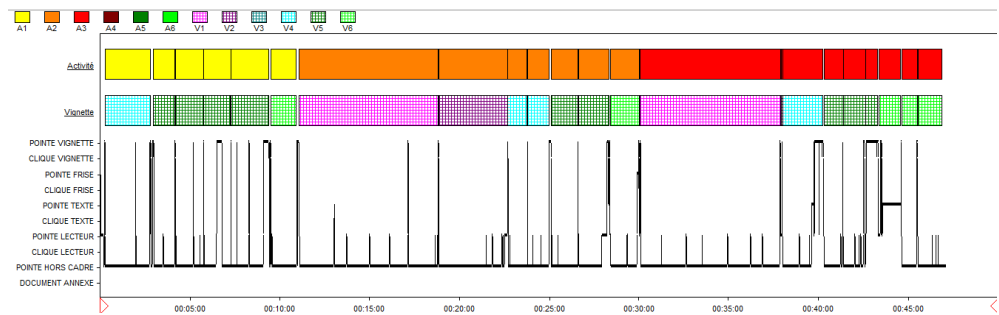

46    Associé au graphe d'organisation intrinsèque (et potentiellement à un traitement statistique), le graphe d'organisation extrinsèque constitue une représentation synthétique heuristique favorisant l'analyse technologique de l'ENF : l'ordre et la nature des ressources consultées ainsi que la succession, la fréquence et la durée des actions produites dans l'interface apparaissent de façon très exploitable. La comparaison entre les graphes est également facilitée. La première ligne correspond à l'activité typique consultée (de 1 à 6) : on peut notamment observer leur ordre de consultation et comparer le temps passé dans chacune d'elles. La seconde ligne correspond au type de vidéo (en classe, vécu professionnel, compléments, débutants, expérimentés, chercheurs) : elle permet le même type d'analyses que la ligne précédente, mais aussi par exemple la mise en relation des types de vidéo et des activités typiques auxquelles ils se réfèrent. On repère ainsi une consultation inégale des débutants selon les activités typiques, un plébiscite des expérimentés de l'activité typique n° 3, un délaissement des chercheurs des activités typiques n° 3, 4, 5 et 6, etc. Les lignes suivantes du graphe correspondent aux types d'actions produites : on peut par exemple constater que la fréquence des actions de pointage, plutôt exploratoires, diminue fortement après les dix premières minutes (cela traduit un changement, voire une « bascule » dans l'engagement de l'acteur).

47      Le choix des couleurs pour chacune des Figures 4 et 5 est arbitraire : elles n'ont pas de lien entre elles.

Figure 5 : Extrait du graphe de description intrinsèque de l'activité d'Anne
*Figure 5: Extract from the intrinsic description graph of Anne's activity*

48      Le graphe d'organisation intrinsèque constitue une représentation synthétique très heuristique pour l'analyse régressive des données : l'identification des invariants et variations est en effet facilitée par la vue d'ensemble qu'il procure des relations de cohérence séquentielles (les cases), macro-séquentielles (les quatre lignes de couleur) et sérielles (les lignes mauves et oranges pour la série 1 et les lignes jaunes et vertes pour la série 2).

## Conclusion

49      Cette étude pilote avait une triple visée : (i) comprendre et modéliser l'organisation locale de l'activité d'un utilisateur (enseignant) en situation d'utilisation de l'ENF, en cherchant des similarités et récurrences dans son cours d'action (ii) dans la perspective de contribuer, dans les études suivantes, à la compréhension de l'activité d'autres utilisateurs par comparaison, assimilation et différenciation des modélisations produites et (iii) afin d'alimenter, *in fine*, un processus de conception continuée par l'identification de leviers possibles d'amélioration de l'ENF. Nous avons montré dans cet article qu'un observatoire du cours d'action d'un utilisateur tel que nous l'avons conçu et mis en œuvre rendait possible la modélisation typologique et chronologique de son activité en unités significatives de différents rangs (unités élémentaires, séquences, macro-séquences, séries). La description extrinsèque et intrinsèque de l'activité d'Anne au cours de sa première utilisation de l'ENF a notamment permis de mettre en exergue quatre engagements-types se manifestant de manière récurrente (voir 5.3). L'étude systématique à suivre, consistant à répliquer le protocole pour chaque enseignant-utilisateur de la cohorte, permettra de mettre en perspective les différents cours d'action pour identifier d'éventuelles similarités autorisant un premier degré de généralisation des résultats (i) dans l'organisation de leurs activités respectives (notamment dans la dimension de l'utilisabilité de l'ENF) et (ii) dans la nature des apprentissages (dimension de l'utilité de l'ENF). Une dernière phase permettra d'envisager une boucle de reconception de l'ENF afin de l'améliorer, notamment sur le plan de la « scénarisation médiatique » (Henri, Compte, & Charlier, 2007) qui préside à la conception de l'ENF, dans la mesure où celle-ci favorisera ou non les apprentissages escomptés.

50    La méthode d'entretien de remise en situation à l'aide de traces de l'activité que nous avons utilisée présente plusieurs avantages vis-à-vis des méthodes de verbalisations simultanées et interruptives. Sur le plan de la description extrinsèque, elle trouve une utilité particulière pour (i) disposer de traces numérique susceptibles de compléter et d'enrichir une prise de notes en temps réel, et permettre une analyse plus fine, éventuellement à l'aide de logiciels dédiés (d'acquisition, de synchronisation, etc.) comme ça a été le cas dans notre étude ; et (ii) différer l'analyse des données qui n'ont pas un intérêt immédiat et qui peuvent être traitées tout ou partie hors du lieu et du temps de réalisation de l'activité (chronométrie, fréquentiel, analyse de postures, de déplacements, de trajectoires, etc.) Ceci peut également permettre au chercheur de se concentrer sur des éléments difficilement « saisissables » et a fortiori enregistrables en situation. Sur le plan de la description intrinsèque, elle trouve une utilité particulière pour différer les interactions avec le chercheur, c'est-à-dire les questions, les relances, les reformulations et les dialogues nécessaires à une intelligibilité partagée. D'une part elles ne sont pas toujours possibles (travail en hauteur, espace confiné, zone à risque, activités sportives, de service, de soin, tâches à forte composante attentionnelle et à fort risque d'erreur, etc.), et d'autre part elles constituent une forme de médiation externe qui n'est pas toujours souhaitée, comme dans notre étude simulant une utilisation autonome d'un ENF.

51    À ce stade, la perspective technologique de notre travail se décline selon deux temporalités. D'abord, à court terme, il s'agit d'étendre la méthode à l'étude des situations d'utilisation de tous les participants à la recherche, afin de (i) construire un corpus robuste et permettre un premier niveau de généralisation des résultats et (ii) définir les aménagements de l'ENF prioritaires et/ou les moins couteux sur le plan technico-organisationnel (relativement aux compétences et ressources mobilisées). Puis, dans une « boucle longue » (Sève, Theureau, Saury, & Haradji, 2012), il s'agit de systématiser les résultats concluants (les expériences vécues par les utilisateurs présentant le plus fort degré de typicité), afin de (i) formaliser de nouveaux principes de formation, applicables à des ensembles de situations débordant celles étudiées et (ii) déterminer la faisabilité et planifier l'effectuation des aménagements de l'ENF les plus couteux sur le plan technico-organisationnel. Cela implique un ensemble de bénéficiaires dépassant celui des seuls participants à la recherche : il s'agit de reconcevoir l'interface de l'ENF étudié, définir de nouveaux principes et méthodes de conception, et informer la conception de nouveaux ENF et dispositifs. L'activité de conception est ainsi envisagée comme une interaction permanente entre compréhension et création des situations présentes en vue de situations futures (Norman, 1986 ; Theureau & Jeffroy, 1994 ; Winograd & Flores, 1989). Dans le cadre de la conception continuée d'un dispositif complexe, comme par exemple l'ENF NéoPass@ction, il serait également utile de définir des boucles d'aménagement-reconception selon différentes temporalités, notamment en lien avec leur coût technico-organisationnel. Un tel outillage conceptuel et méthodologique permettrait d'envisager plus efficacement de nouveaux projets de conception itérative, notamment d'environnements numériques.

---

## BIBLIOGRAPHIE

## NOTES

**1.** NéoPass@ction, une plateforme de formation de l'Institut Français de l'Éducation - ENS de Lyon : http://neo.ens-lyon.fr

**2.** Application *FastStone Capture :* http://www.faststone.org - consulté le 19 janvier 2016.

**3.** « Lorsque l'engagement est exploratoire, les préoccupations de l'acteur sont relatives à la compréhension de la situation. Il fait preuve d'une intense activité interprétative, et agit de façon à vérifier ou construire de nouvelles connaissances. Lorsque l'engagement est exécutoire, l'acteur recherche une efficacité maximale. Il s'appuie sur la familiarité des situations rencontrées pour apporter une réponse déjà éprouvée dans des situations similaires. » (Sève &

Leblanc, 2003, p. 65). Ces catégories sont proches de celles proposées précédemment par Norman (1986), dans une approche mentaliste, en termes d'exécution et d'évaluation.

**4.** Cela ne s'est jamais produit pour cette étude. L'utilisabilité de l'ENF s'est révélée suffisante pour rendre possibles toutes les actions initiées par l'utilisatrice.

**5.** « Scroller » signifie faire défiler verticalement le contenu d'un document sur un écran d'ordinateur à l'aide de la molette d'une souris, d'un pavé tactile ou d'un curseur graphique interactif (souvent une barre latérale à l'extrémité droite de la fenêtre).

**6.** *Kinovéa*, logiciel de lecture vidéo : http://www.kinovea.org, consulté le 25 janvier 2016.

**7.** *Actogram*, logiciel de traitement de relevés horodatés : http://www.preventica.com/doc-actogram-kronos-octares-editions.php, consulté le 25 janvier 2016.

**8.** Dans une approche enactive telle que la nôtre, on postule la cohérence de l'activité (pour tout organisme sain), entendue comme auto-organisation « équilibrante », sans considération logique. On s'intéresse donc aux différents niveaux de cohérence de l'activité et aux différentes cohérences possibles.

---

## RÉSUMÉS

Cet article est une contribution empirique au domaine de la formation instrumentée par le numérique mais aussi — et surtout — une contribution méthodologique à l'analyse des activités produites dans ce domaine. Il rend compte d'une étude pilote menée dans le cadre d'une recherche doctorale et qui a consisté à décrire, analyser et modéliser l'activité d'un enseignant stagiaire en situation d'utilisation autonome d'un environnement numérique de formation (ENF) basé sur des extraits vidéo. Un soin particulier est apporté à la description de la méthode. Deux types de données ont été recueillis et traités selon l'approche du programme « cours d'action » : (i) des données d'observation de l'activité (capture dynamique d'écran) et (ii) des données d'entretien de remise en situation à l'aide des traces numériques de cette activité. Les résultats (i) valident la pertinence de la méthode vis-à-vis de l'objet et de la question de recherche ; (ii) montrent des structures de différents niveaux dans l'organisation de l'activité en situation d'utilisation ; (iii) révèlent quatre registres de préoccupations (ou engagements-type) orientant l'utilisation de l'ENF. Nous concluons dans une perspective de technologie de la formation, en discutant la façon dont ces résultats informent, à certaines conditions et selon différentes temporalités, un processus de conception continuée de l'ENF.

This article is an empirical contribution to the field of educational technology but also — and above all — a methodological contribution to the analysis of the activities enacted in this field. It takes account of a pilot study conducted within the framework of doctoral research and consisted in describing, analysing and modelling the activity of a trainee teacher in a situation of autonomous use of a video-based digital learning environment (DLE). We were particularly careful to describe the method in great detail. Two types of data were collected and processed within the framework of "course-of-action": (i) activity observation data (dynamic screen capture) and (ii) data from resituating interviews supported by digital traces of that activity. The findings (i) validate the method's relevance in relation to the object and issues of the research; (ii) show different levels of organization in the activity deployed in the situation of use; (iii) highlight four registers of concerns orienting use of the DLE. We conclude from a perspective of educational technology, by discussing how, according to certain conditions and different time scales, the findings inform a process of continuous DLE design.

## INDEX

**Mots-clés :** environnement numérique de formation, entretien de remise en situation à l'aide de traces de l'activité, cours d'action, formation des enseignants
**Keywords :** digital learning environment, video-enhanced education, resituating interviews supported by activity traces, course-of-action, teacher education

## AUTEURS

**SIMON FLANDIN**

Faculté de Psychologie et de Sciences de l'éducation, Laboratoire RIFT, Équipe CRAFT
Université de Genève, 40 boulevard du Pont d'Arve - CH 1211 Genève 4
simon.flandin@unige.ch

**MARINE AUBY**

E.T. ergonomie, 6 place du 8 mai 1945 - 69670 Vaugneray
marineauby@hotmail.fr

**LUC RIA**

Laboratoire ACTÉ (EA 4281)
Institut Français de l'Éducation, ENS de Lyon, 15 parvis René Descartes 69347 Lyon Cedex 07
luc.ria@ens-lyon.fr